\documentclass[emulateapj,epsfig]{article} 
\usepackage[onecolumn]{emulateapj}
\usepackage[]{epsfig} 
\usepackage{graphicx}
\hoffset -1.5truecm \lefthead{Menci et al.} 
\righthead{} 
\def\newline{\hfil\break} 
\begin{document} 
\title{Bimodal AGNs in Bimodal Galaxies } 
\author{A. Cavaliere$^1$, N.  Menci$^2$} 
\affil{$^1$Dip. Fisica Univ.  di Roma ``Tor Vergata'', via  Ricerca Scientifica 1, 00133 Roma, Italy}
\affil{$^2$INAF - Osservatorio Astronomico di Roma, via di Frascati 33, 00040 Monteporzio, Italy}
\smallskip 

{\begin{abstract}
By their star content, the galaxies  split out  into a red and a blue 
population; their  color index peaked  around $u-r\approx 2.5$ or  $u-r\approx 
1$, respectively, quantifies the ratio of the blue stars newly formed  from cold 
galactic gas,  to the redder ones left over by past generations. On the other hand, upon accreting 
substantial gas amounts the central massive black holes energize active galactic nuclei (AGNs); 
here we investigate whether these show a similar, and possibly related, bimodal 
partition as for current accretion activity relative to the past. To this aim we use  
an updated semianalytic model; based on Monte Carlo 
simulations, this follows with a large statistics the galaxy assemblage, the 
star generations and the black hole accretions in the cosmological framework 
over the redshift span from $z=10$ to $z=0$.
We test our simulations for yielding in close detail the observed split of 
galaxies into a red, early and a blue, late population. We find that the 
black hole accretion activities likewise give rise to two source populations: 
early, bright quasars and later, dimmer AGNs.  We predict for their Eddington 
parameter $\lambda_E$ -- the ratio of the current to the past black hole accretions -- a 
bimodal distribution; the two branches sit now under $\lambda_E \approx 0.01$ 
(mainly contributed 
by low-luminosity AGNs) and around $\lambda_E \approx 0.3-1$. 
These not only mark out the two populations of AGNs, but 
also will turn out to correlate strongly with the red or blue color of their host galaxies.

\end{abstract}


\section{Introduction}

With the beginning of the 2000s the 
galaxies -- long known to differ by articulated morphologies (see Sandage 2005) -- 
have been recognized   
to neatly split by the colors of their stellar content  into two  populations 
of red or blue objects (Strateva et al. 2001, Baldry et al. 2004).  

The {\it red} population comprises giant spheroidal 
galaxies larger than $M\approx 10^{12}\,M_{\odot}$. These are found to be in place 
by redshifts $z\approx 2$ when the Universe was only a fifth of its present age; thence 
they evolved ``passively" by declining stellar birthrates. So they appear as 
precociously ``red and dead", marked  by low values around $0.1$ of the ratio 
$\eta = L_U/L_R$ 
of the UV-blue to the red luminosity, that is,  by a color index $u-r = -2.5\;  log\,\eta \approx 2.5$ 
(AB photometric system). The {\it blue} population, instead, is mainly comprised 
of smaller galaxies featuring a spheroidal bulge but also a conspicuous disk, 
with star formation still lively today  so as to retain values $\eta\approx 1$ 
($u-r \approx 1$). 

These two populations correlate with their environment, with the former 
mainly inhabiting dense, crowded sites like groups or clusters of galaxies, and 
the latter  widely distributed in the ``field". The color bimodality is being 
traced 
(Bell et al. 2005, Giallongo et al. 2006) into the deep Universe out to $z\approx 2$ when the galactic morphologies 
were mixed or clumpy (see Elmegreen et al. 2007).

The other component, harboured at the very center of most bright galaxies, is a 
massive  black hole (BH, see Ferrarese \& Ford 2006)  with  mass in the range $M_{\bullet} \approx 10^6-10^9 \, M_{\odot}$. 
 This  is some $10^{-3}$  of  the mass in the host spheroidal bulge; but on accreting a comparable mass at 
rates up to $\dot M_{\bullet} \approx 10^2\, M_{\odot}$/yr converted to energy at 10 \% efficiency, these BHs 
easily outshine all the galactic  stars as  active galactic  nuclei (AGNs) or quasars radiating for some  $10^8$ yr bolometric 
luminosities up to $L \approx 0.1\, c^2\, \dot M_{\bullet} \approx 10^{14} L_{\odot}$.  When this exceeds the Eddington luminosity 
$L_E = 3\, 10^4\,  M_{\bullet} \; L_{\odot}/M_{\odot}$ the very radiation pressure can blow away a surrounding source, 
whereas at $L = L_E$  the mass $M_{\bullet}$  is minimized. So the bright, early quasars 
observed (Fan et al. 2004) to radiate about $L \sim  10^{14}L_{\odot}$ at $z \approx 6$ must conservatively involve 
supermassive BHs with $M\approx 10^9 M_{\odot}$ activated at Eddington levels $L/L_E \approx 1$.  Less 
clear conditions prevail in later and fainter AGNs; these we discuss next, with 
focus on predicting the Eddington ratios 
$\lambda_E = L/L_E \propto \dot M_{\bullet}/M_{\bullet}$ 
that the observations estimate with still significant selections and uncertainties.

\section{Semi-analytic modeling}

We calculate the 
distributions of BHs and host galaxies together on using a single semi-analytic 
model (SAM); this links several astrophysical processes to yield the 
astronomical observables with a large statistics within manageable computer 
times. 

The building blocks are provided by Monte Carlo simulations 
for the growth of the dark matter condensations (DM ``halos") that dominate the 
mass in all cosmic structures from galaxies to their groups and clusters. As the 
Universe ages, expands and thins down, the primordial density fluctuations 
condense out on progressively larger scales by gravitational instability (cf. Peebles 1993). 
Their growth actually involves major mergers with comparable halos and coalescence with 
smaller ones; after each round  a halo goes to a virial equilibrium with internal 
densities $\rho \approx 10^2 \, \rho_u(z)$ relative to its surroundings. 
Details of our computations are given in the Appendix. 
The building up of a large present-day galaxy through the above dynamical 
``merging histories" is illustrated in Fig. 1. 
 
\begin{center}
\vspace{0.5cm} 
\scalebox{0.45}[0.45]{\rotatebox{0}{\includegraphics{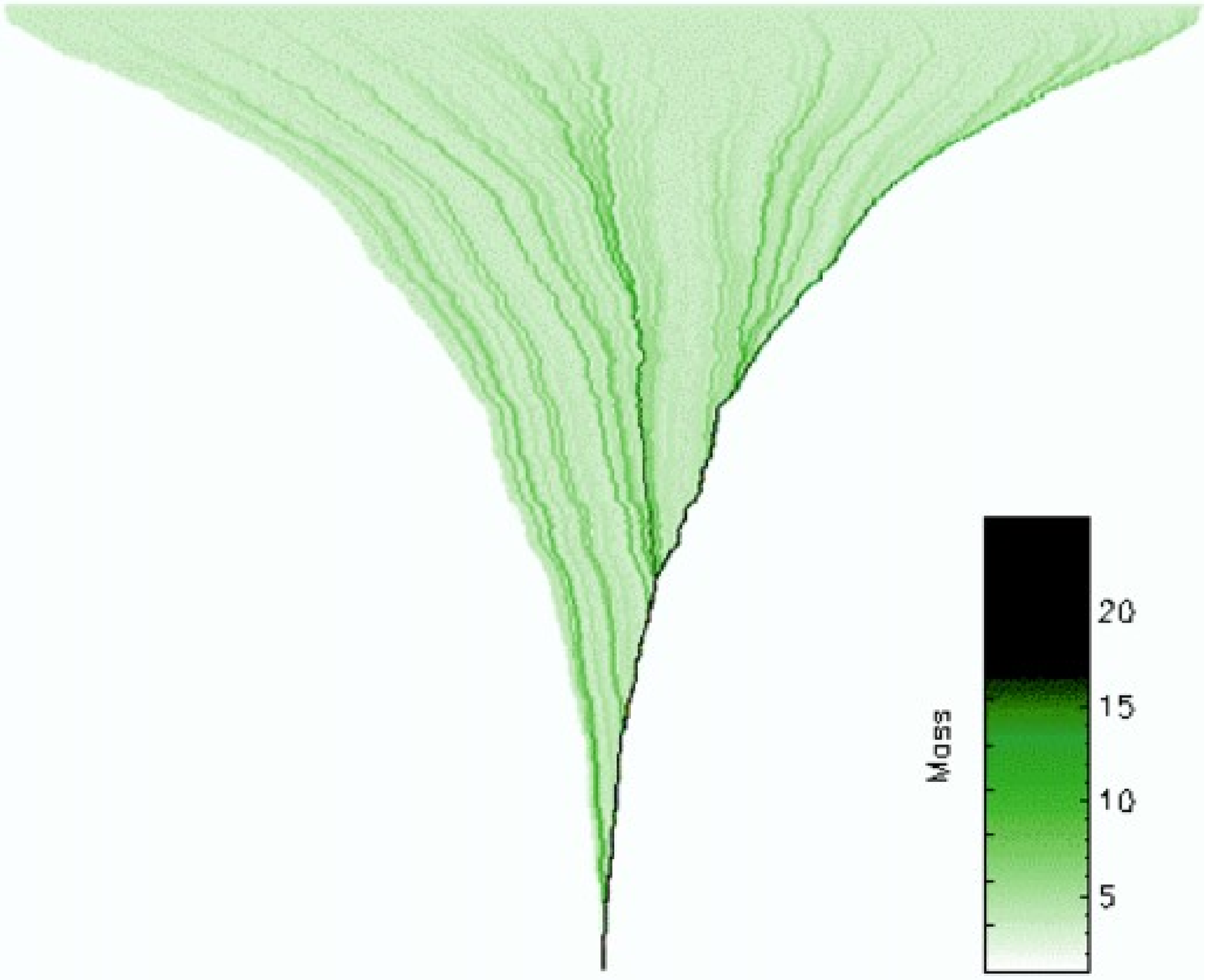}}}
\end{center}
{\footnotesize
\vspace{0cm } Fig. 1. - 
A merging tree from our Monte Carlo simulation, showing the formation history of 
a central dominant galaxy in a large DM halo with present mass $10^{13}\, M_{\odot}$. 
Time runs from top to bottom, from $z=10$ to the present. Each branch represents 
a progenitor of the final galaxy, and the color code (shown on the right in units of
of $5\,  10^{11}\,  M_{\odot}$) quantifies the mass 
of the corresponding progenitor. 
A total of some $10^3$ progenitors are involved, 
with the main one being represented on the rightmost branch.  
\vspace{0.3cm}}

As the DM halos are included into larger ones they may survive as 
sub-halos holding galaxies; these, however, may coalesce at the center to form a 
dominant galaxy, or orbit for a while within the halo as satellite galaxies. In our 
simulations we compute in detail 
the dynamical processes causing evolution of these galaxies, namely: 
i) the dynamical friction causing inspiral and 
leading to coalescence with the central galaxy; ii) the binary aggregations 
between two satellite galaxies orbiting within a common DM halo;
iii) other grazing encounters of  galaxies not leading to merging, but
affecting star formation and BH accretion (see  below).

As for the baryonic component, 
the dark halos provide gravitational wells for the visible galaxies 
to sit in and cycle baryons at densities large enough to form stars and grow BHs, 
as computed by  SAMs (see Appendix). Briefly, stars form quiescently at rates $1-
10\,  M_{\odot}$/yr after the classic Schmidt birthrate $\tau_*^{-1}\propto m_c/t_d$ in terms of the 
dynamical time scale $t_d$ and of the mass $m_c$ of cold galactic gas. The latter  is 
modulated to a sensitive function of time $m_c(t)$ by a number of standard 
processes: consumption by the very star formation;  depletion upon outflows caused by  
the energy fed back from Supernova explosions ending the life of massive  stars; 
additions from initially virialized but radiatively  cooled  gas. 

In addition, our SAM also includes impulsive star formation up to several 
$10^2 \, M_{\odot}/$yr as occur in a 
protogalaxy when a collision with a companion stimulates a burst rich in 
blue-UV stars.  From major mergers to minor merging to close fly-byes, these gravitational interactions 
distort the gravitational field and cause  some galactic gas to lose angular 
momentum; thus inflow and convergence are  stimulated, and stars form in plenty at 
effective birthrates (see Appendix, Subsect. "Collisions and interactions") 
$\tau^{-1} = N\,\Sigma\, V   \propto \rho(z)\,M^{2/3}$, 
a scaling that favours large galaxies  at 
high $z$. A related feature of our SAM is the gas fraction  from the same inflow 
that  reaches further down  to the very center to fuel a massive BH and kindle 
quasar emissions (Cavaliere \& Vittorini 2000, 2002; Menci et al. 2003; 
Hopkins et al. 2006). 
These in turn cause violent energy feedback on the 
surrounding gas, to the point of halting accretion or even quenching all star 
formation, see also Di Matteo, Springel \& Hernquist (2005). 

Linking and bringing to focus all these processes, our SAM yields galactic scaling 
laws and luminosity functions in agreement with the data at low and high $z$ 
(see Menci et al. 2003, 2005). Next we highlight additional results and their interpretation.  

\section{Bimodal galaxies}

In Fig. 2 we show our galaxy color distribution at $z \approx 0$, which  
splits out into a neatly {\it bimodal} population in detailed agreement with the 
observations; the lower panels articulate our  projections to,  and predictions at higher $z$. 

\begin{center}
\vspace{-8cm} 
\scalebox{0.78}[0.78]{\rotatebox{0}{\includegraphics{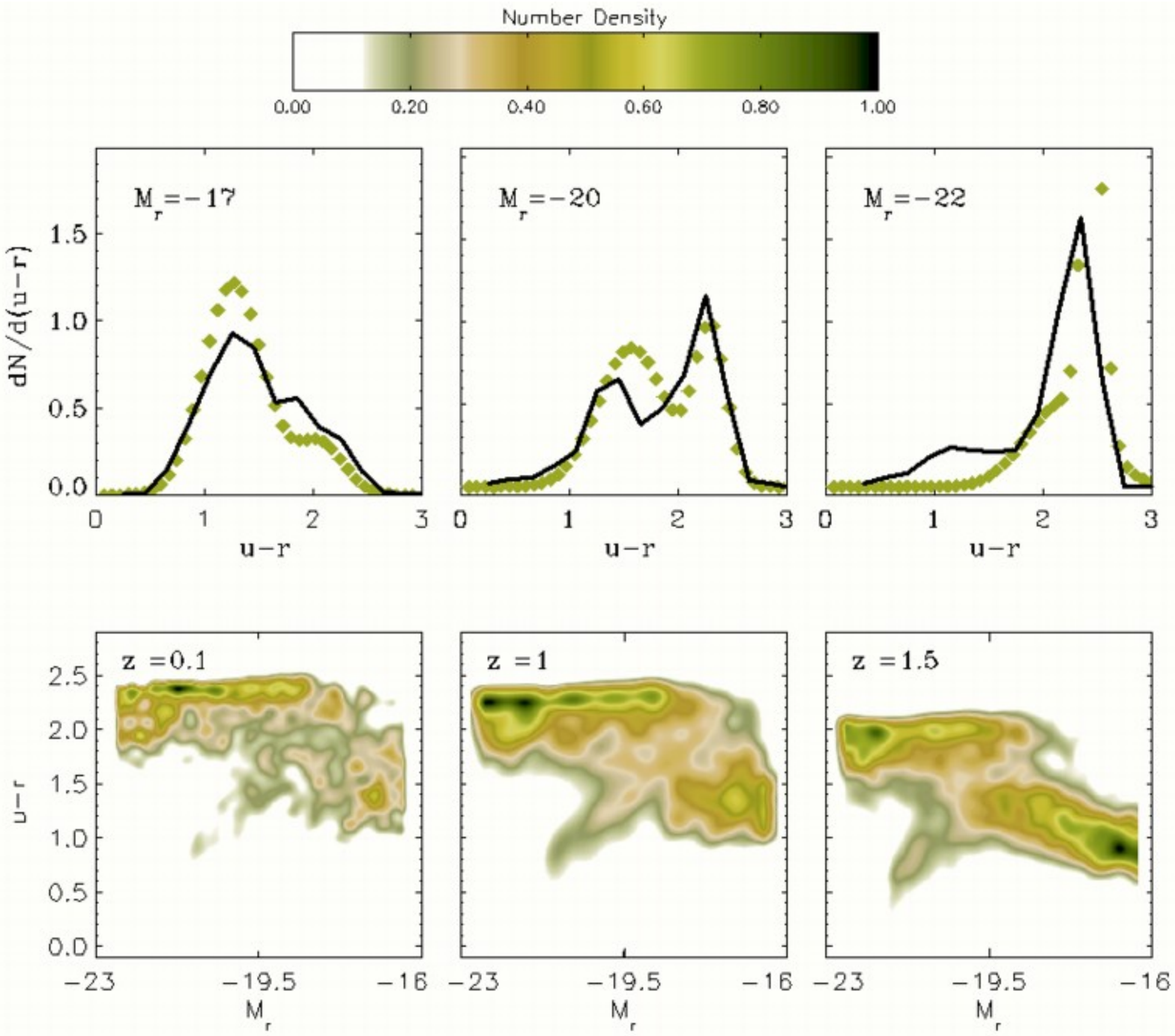}}}
\end{center}
{\footnotesize
\vspace{0cm } 
Fig. 2. - Bimodal galaxies from our SAM.  {\it Upper panels}:  plots of the present ($ z = 
0$) color distribution in the form of  number density $dN/d(u-r)$ normalized to 
its maximum  vs. the color index $u-r$, at different absolute magnitudes $M_r$ 
in the red band.   Clear bimodality appears as the  bright population marked by 
{\bf red} colors  $u-r \approx  2.5$  ($\eta \approx 0.1$) peaks away  from  the 
fainter, {\bf bluer} one  ($u-r \approx 1$, $\eta \approx 0.4$), and compares 
well with the extensive  observational statistics (diamonds, see Baldry et al. 
2004). {\it Lower panels}:  contour representation of the evolving color-
magnitude relation that we predict at higher  $z$.  Bimodality persists  out to 
$z = 1.5$, with  the ``red sequence"  still separated  by  a   ``valley" from 
the  ``blue cloud";  while the latter is progressively stretched out into  a 
richer   ``blue sequence",   the former is 
progressively skewed toward the bright  massive end.

\vspace{0.3cm}}

Clearly we are confronted with two effectively  {\it different} chains of events started by the DM 
merging histories and amplified by the star formation processes, that we 
reconstruct as follows. 

The growth of a halo is speeded up when its initial seed 
is biased high, i.e., grows in an {\it overdense} and {\it crowded} region of the primordial 
Gaussian fluctuations (Bardeen et al. 1986).  The associated 
protogalaxies undergo more frequent mergers and other gravitational interactions (Conselice 2006) in the early, denser Universe; thus  they 
build and light up precociously, and  assemble into the large spheroidal galaxies 
with $M >10^{12} M_{\odot}$ of today. Mergers occurring at protogalactic scales also replenish 
the gas masses  $m_c$  toward the cosmic ratio  of baryons/DM $\approx 0.17$; so the 
stellar birthrates are sustained up to several $10^2\, M_{\odot}$/yr in stimulated bursts. 

But when (by $z < 2$) major galaxy merging becomes increasingly rare, the gas 
reservoirs rapidly run out and all stellar birthrates run down. Then the early 
era of efficient star formation comes to an end, leaving behind a multitude of 
reddish, sluggish stars that taint these large  protogalaxies already  early on 
while they assemble into the giant, red and dead  galaxies of today.  

Meanwhile, smaller galaxies assemble from lower-mass progenitors growing up in 
the more common, {\it lower-density} and {\it quieter} regions of the primordial 
fluctuation field, and  quiescently process their gas into stars at a lower but 
more even pace. So  this other branch of histories leads to galaxies that retain 
larger amounts of cold gas, conducive to enduring if peaceful stellar birthrates 
and regeneration of massive stars with blue color. 
  
Even these  galaxies may make a transition to the fully inefficient star-foming 
regime and turn red. This occurs after they undergo a late major merger 
(Conselice 2006) that triggers a  gas-consuming  starburst, or when they 
suddenly grow to a size such that gas cooling is inefficient (Dekel \& Birnboim 
2006), or in the long run eventually by sheer  exhaustion of their cold gas 
reservoir. Overall, at cosmic ages later than $z \approx 2$ the actual star 
formation activity appears as if ``downsized" toward smaller galaxies (Cowie et 
al. 1996).

Such specific evolutionary behaviors begin to be captured by recent models of 
galaxy formation (Menci et al. 2005; Bower et al. 2006; Cattaneo et al. 2006; 
Croton et al. 2006), and are focused by the present one in connection with the 
AGN behavior.  

\section{Bimodal AGNs}

Here we stress a related bimodality in AGN activity, that we expect  since the 
galactic gas feeds both the star formation and the central BH accretion.  Our 
SAM specifically links the bolometric AGN luminosity $L \propto f m_c$ to the 
cold gas amount $m_c(t)$ left over at the time $t$ and to its fraction $f$ 
stimulated to accrete; this  is statistically distributed since the interaction 
parameters are set by the  merging history specific to each host galaxy. The 
(possibly repeated) accretion events accrue the summed mass $M_{\bullet} = \int 
dt L(t)/ 0.1 \, c^2$ in the BHs that constitute the endpoints of gravitational 
collapse.  

\begin{center}
\vspace{-4cm} 
\scalebox{0.61}[0.61]{\rotatebox{0}{\includegraphics{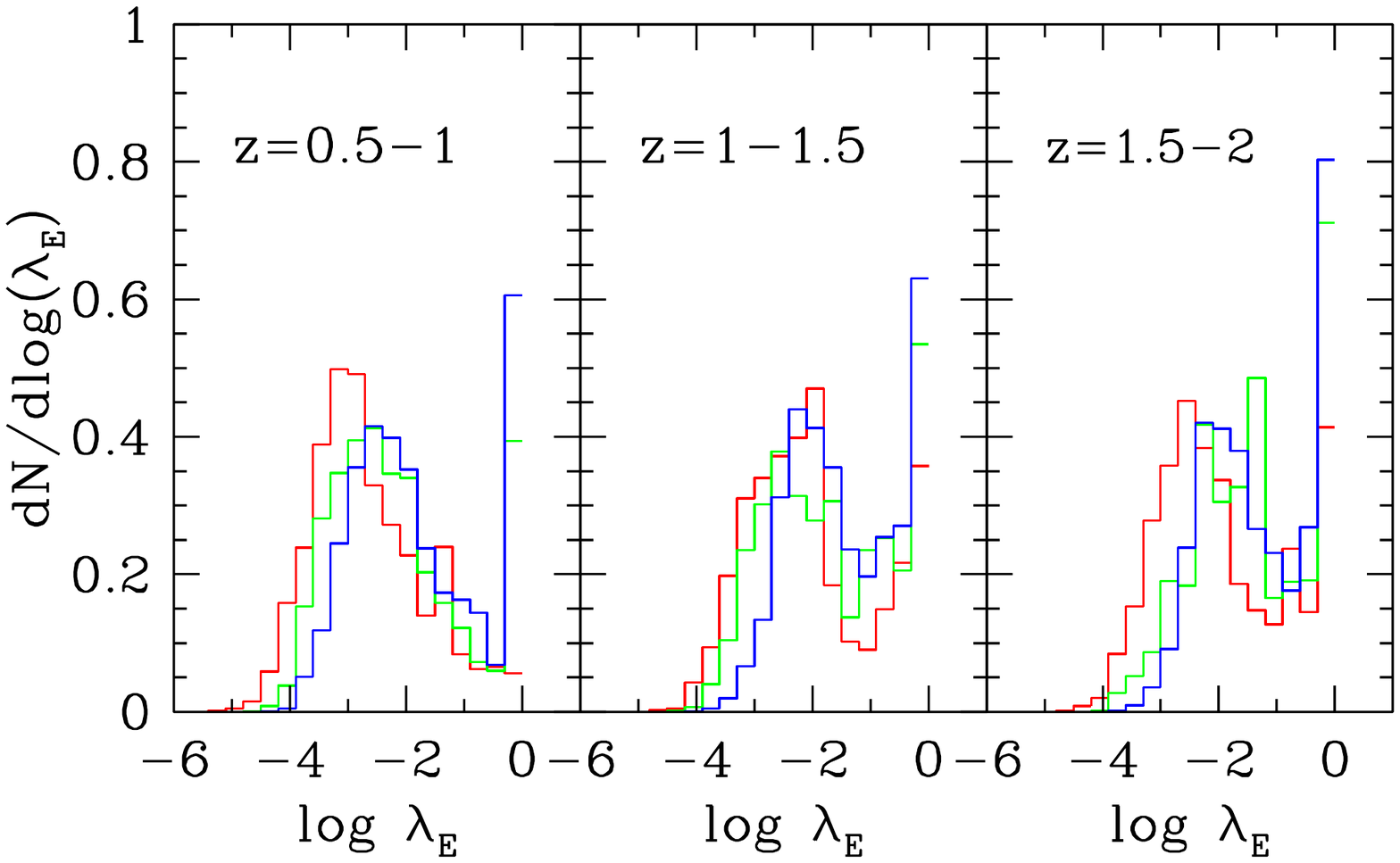}}}
\end{center}
{\footnotesize
\vspace{-4.cm } 
Fig. 3. - Bimodal AGNs from our SAM. 
In three redshift bins we show the distributions of $\lambda_E$ for AGNs with different luminosities: 
$L>10^{45}$  (blue histogram), $10^{44.5}\leq  L \leq 10^{45}$ (green histogram), 
$L< 10^{44.5}$ (red histogram); the distributions are normalized to the total content of each luminosity range. 
\vspace{0.3cm}}

From $L$ and $M_{\bullet}$ we derive our predictions concerning the 
distributions of the Eddington parameters $\lambda_E = L/L_E$ and show them in 
Figs. 3 and 4; we find  the AGNs to exhibit a {\it bimodal} distribution of 
$\lambda_E \propto \dot M_{\bullet}/M_{\bullet}$ that quantifies the current 
relative to the past BH accretion. The content of the two branches changes with 
redshift; specifically, in the era of intense star formation for $z > 1.5$  most 
AGNs stay up to $\lambda_E \approx 0.3-1$, as observed (see McLure \& Dunlop 
2004, Vestergaard 2004), with the brighter AGNs skewed toward larger values of 
$\lambda_E$.
At lower $z$, instead,  we predict the branch with $\lambda\sim 10^{-2}$ to 
become increasingly richer mainly in low-luminosity ($L\lesssim 10^{44}$ erg/s) 
AGNs,  in accord with observations by Panessa et al. (2006) and findings by Lapi 
et al. (2006). We specifically predict (see Fig. 4) that at low redshifts most 
AGNs with very small $\lambda_E$ are hosted in {\it red} galaxies, whilst those 
with larger values around $\lambda_E \approx 0.3-1$ are mainly hosted by {\it 
blue} galaxies. 

This is because protogalaxies on the early, hectic branch of the merging 
histories process much of their gas content both into vigorous star formation, 
and into bright quasar  activity sustained at full levels $\lambda_E \approx 1$ by frequent mergers and plenty of gas. 
But soon after $z\approx 2.5$ these activities concur to deplete 
the gas, and drive themselves into a steep decline. What remains of those 
flaming days - along with the multitude of  red stars that taint such galaxies -
are dozing or dormant supermassive BHs with trickling, inefficient accretion. So we 
expect the early quasar population that shone up at high z in dense environments 
to fall down in luminosity while retaining their large accrued BH masses; these 
features concur to confine the Eddington ratios to small values $\lambda_E \lesssim 10^{-2}$ 
in red  hosts at low z. 

Along the other, easy-going branch,  smaller and gas-rich 
galaxies not only feed quiescent but persisting star formation and regenerate 
blue stars, but can also provide to smaller BHs material for sizeable accretion 
energizing a late AGN population (cf. observations by Kauffmann et al. 2003, Vanden Berk et al. 2006), when effectively 
stimulated. A main trigger is again provided by galaxy interactions, rare but 
persisting in the ``field"; these cause gas inflows in the host, either directly 
or  accruing to  yield  secular disk instabilities (see Combes 2005).  The basic rate 
$\tau^{-1} \propto\rho$ (Cavaliere \& Vittorini 2000, and Appendix) takes on values $\tau^{-1} \approx 0.1$/Gyr 
in the Large Scale 
Structures that streak the ``field" with galaxy density contrasts   $\rho/\rho_u \approx 3-5$, 
as  compared to $\tau^{-1}\gtrsim 1$/Gyr  in early small groups some 10 times denser. 

These field interactions act on  a  gas content generally rich but variable in 
detail. Thus we expect for the late AGN population values of $\lambda_E$ considerably {\it exceeding} those 
in the red hosts at equal luminosities, but {\it scattered} in a wide distribution 
that can be accurately  predicted only by a numerical code such as our SAM (see 
Figs. 3 and 4). On similar grounds one expects a wide scatter also for the values of $\eta$  
in the blue galaxy population, as in fact we had found and shown in Fig. 2. 

\begin{center}
\vspace{0cm} 
\scalebox{0.6}[0.6]{\rotatebox{0}{\includegraphics{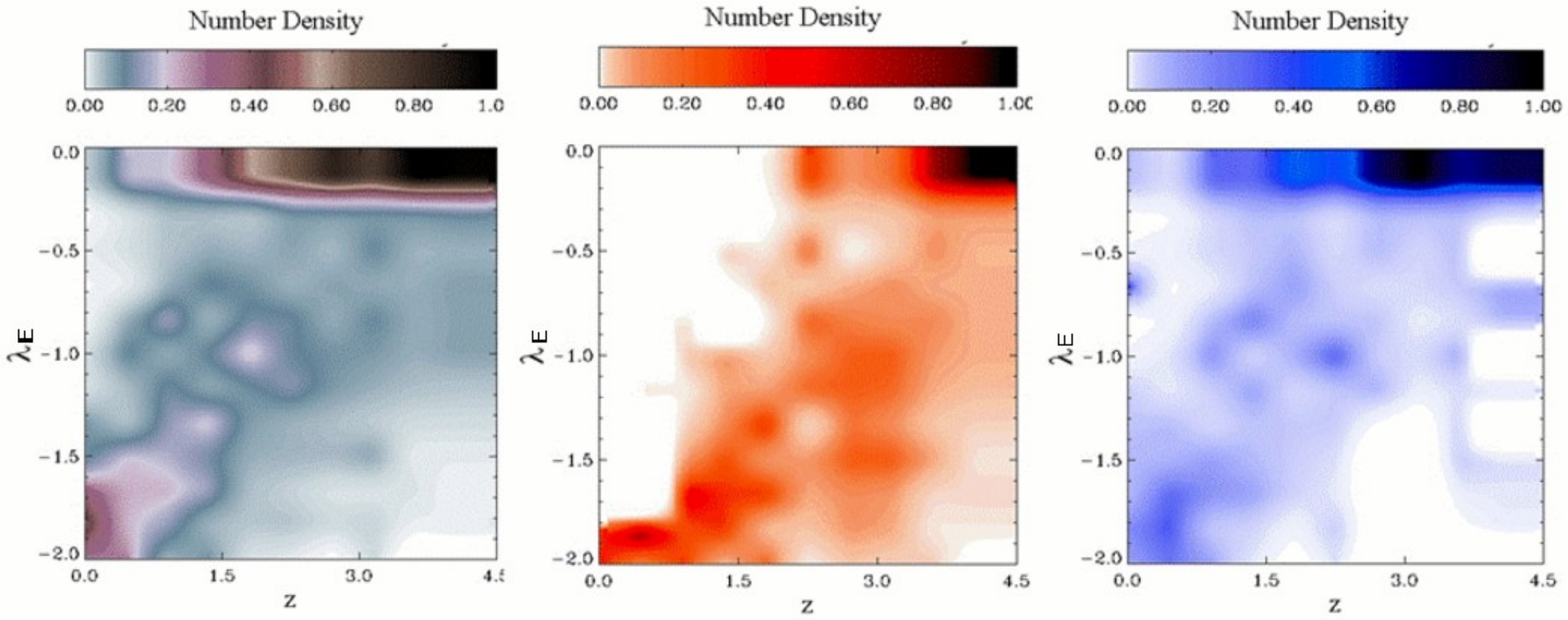}}}
\end{center}
{\footnotesize
\vspace{-6.cm } 
Fig. 4. - 
Bimodal AGNs from our SAM. 
{\it Left}: the overall evolution with $z$ of the 
Eddington ratios $\lambda_E$
for AGNs brighter than $L=  10^9\, L_{\odot}$; the code represents the number density 
of galaxies, normalized to the 
maximum value at each $z$. 
{\it Middle}: same, for the AGNs hosted in {\bf red} galaxies ($u-r >  1.5$). {\it Right}: 
same, for AGNs in {\bf blue} galaxies ($u-r \leq  1$).
}

\section{Conclusions and discussion}

In sum, physical evaluations borne out by our advanced SAM yield the articulated 
demographics illustrated by Figs. 2, 3 and 4, and lead us to the following picture. 
The evolutions of the star glow and of the quasar-AGN shine may be collectively 
described as ``downsizing" of all activities, but these are specifically marked by 
{\it bimodal} and {\it corresponding} ratios of current/past star generations ($\eta =  L_U/L_R$) 
and current/past  BH accretions ($\lambda_E \propto L/M_{\bullet}$). In fact, all 
baryon
 emissions are driven along 
two different evolutionary tracks by the DM merging histories (see Fig. 1); the difference  is 
switched on as the tracks start from biased or from average regions in the 
fluctuation field, and is amplified by the baryonic histories.

Galaxies on biased tracks soon grow large,  while  interacting at brisk rates $\tau^{-1}\propto 
\rho(z)\,M^{2/3}$ that stimulate {\it frantic} burts of 
star formation and BH accretion. Both activities live 
fast and die young by $z\approx 2$, when major mergers run down and galactic gas runs 
out; what remains of those flaming days are large masses stocked in red stars 
and in supermassive BHs. Along the average tracks, instead, all activities proceed 
at an {\it easier} pace that makes up lesser galaxies and BHs, but conserves gas for 
later use. 
These residual cold gas masses $m_c(t)$ can feed {\it persisting} star birthrates
$\tau^{-1}\propto m_c$ that renew blue stars and  bluish overall colors (see Fig. 2).

Meanwhile, the early 
and the late quasar-AGN populations differ even more. This is because red, gas-poor 
hosts can only feed fully grown BHs to yield feeble luminosities $L \propto m_c(t)$, 
upon ineffective interactions in rich groups and clusters. Blue gas-rich hosts, 
instead, can supply to still underdeveloped  BHs sizeable gas lumps to kindle 
them up, when stimulated by external events of {\it interaction} at rates $\tau^{-1}\propto \rho(z)$  slowly 
dwindling  in the field. 

In spite of the 
scatter imparted to the later AGN population by the increasingly sporadic 
fueling of their BHs, such a difference  can be recognized (see Fig. 4) in the distribution 
of $\lambda_E$ when this will be statistically correlated with host color $u-r$. 
 
\bigskip
\noindent
{\bf Acknowledgements}. We thank our referee for comments useful toward tuning up our presentation. 

\bigskip\bigskip\bigskip

\newpage
\centerline {APPENDIX}  
\bigskip

\noindent
{\bf Framework}. We adopt  the 
flat Concordance Cosmology  (see Spergel et al. 2006)  with Hubble expansion constant $H_0=70$ 
km/s Mpc; the matter density with parameter $\Omega_0= 0.3$ is dominated by the DM with a 
baryonic fraction 0.17.

Within this framework, the galaxy evolution is widely described by semi-analytic 
modeling (SAM, e.g., Somerville, Primack \& Faber 2001, Bower et al. 2006). Our 
SAM includes, along with the standard processes, a number of novel features; 
key points  follow, with details given in  Menci et al. (2003, 
2005).  

\noindent
{\bf Dynamical histories}.
Our computations are started at $z = 10$ and cover the range of cosmic times 
down to $z = 0$ in 300 evenly spaced time steps. We take from Lacey \& Cole 1993  the  probability 
for a halo mass $M$ condensed  at a given time $t$ to be merged into a larger mass $M 
+ \Delta M$ at the time $t + \Delta t$, consistent with the classic Press \& Schechter overall 
mass distribution. We run 5,000 merging histories yielding at $z = 0$ a grid of 50 
final halo masses with evenly spaced logarithmic values in the range $10^9-10^{15}
\,M_{\odot}$; for each we compute 100 realizations of merging histories.  

The galaxy histories, started out by assigning a galaxy to each halo, soon take on 
a  course of their own. As {\it t} proceeds and two halos merge, the embedded 
satellite galaxies  
may either coalesce with the dominant one by fast orbital decay
due to dynamical friction, 
or engage in a binary ``collision" with  another satellite. 
When starting in a biased region, a history is 
speeded up; it is easily checked thst galaxy-size fluctuations growing in a group-sized one halve their 
collapse and quarter their interaction times.  

\noindent
{\bf Gas cooling}. 
Stars and black holes feed on cool gas. Initially the gas content 
of a halo is $m_c  = 0.17\, M$, with a temperature at the virial value. The fraction that 
cools radiatively by line and continuous emissions is computed with standard 
procedures described in detail in Somerville et al. (2000, 2001) and Menci et al. (2005). 
The cooled gas is assigned to a disk (in keeping with 
the  observations by Genzel et al. 2006, and by Elmegreen et al. 2007) 
in the dominant galaxy; radius 
$r_d\approx 2$  kpc and rotation velocity 
$v_d\approx 100$ km/s are computed after Mo, Mao \& White (1998), 
and specify the  dynamical  timescale to $t_d = r_d/v_d \approx 10^8$  yr.  
The disk is destroyed and the stars are assigned to a bulge when a merging event involves 
a partner more massive than 1/3 of the main progenitor galaxy (major merger). 
A disk surrounding the bulge may form again when new gas in the halo is able to 
cool anew; this standard procedure is described in detail in Cole et al. 2000.

\noindent
{\bf Quiescent star formation}. 
From the cooled gas stars form at the Schmidt rate $dm_*/dt = 0.1 \,m_c/t_d$ ,  to 
yield $1 - 10 \, M_{\odot}$/yr. 

\noindent
{\bf Supernova feedback}. 
Stars exceeding $8 \,M_{\odot}$ end up  in a Supernova explosion that releases an energy 
$E_S = 10^{51}$ ergs. This re-heates to the hot phase a gas mass $m_h = 3\, 10^{-4}\,E_S\, m_*/v_c^2$ 
in halos with circular velocity $v_c$.  In intermediate-small galaxies with $v_d < 
120$ km/s (see also Dekel \& Silk 1986; 
Dekel \& Birnboim 2006) much cold gas is heated up or even pushed out, so as to enforce self-regulating star formation in  a push-pull regime.  

In addition to the above standard processes our SAM includes the following ones.   

\noindent
{\bf Collisions and interactions}. 
Our SAM includes merging and coalescence of galaxies within common DM halos, due 
to both orbital decay from dynamical friction, and to random collisions with 
other satellite galaxies; the interplay and competition bewteen the two 
processes in dense environments is discussed and illustrated in Menci et al. 
(2002 and references therein). The timescale for dynamical friction and binary 
merging, and so the probability for such processes to occur in each timestep, 
are given by Eqs. (2) and (4) in Menci et al. (2002).
We also compute the probability for a galaxy to undergo a grazing interaction with another 
satellite galaxy in a common DM halo. 
In each halo we compute the rate 
 $\tau^{-1}= N\,\Sigma\, V$ of galaxy interactions from the 
current  number density $N =\rho/M$  of galaxies with average relative velocity $V$ 
and cross section $\Sigma$.  This  is close to  the geometric value, to imply  the  scaling  
$\tau^{-1}\propto \rho\,M^{2/3}$,  for the velocities $V \lesssim 2 v_c$ as occur 
in small groups or in the field; on the other hand,  $\Sigma$ is quite smaller at the higher  $V$ occurring in 
rich groups or clusters. 

{\bf Stimulated gas inflow}. 
Grazing interactions  trigger starbursts and BH accretion (for relevant recent 
observations see Borne et al. 2000; 
Komossa et al. 2003; Guainazzi et al. 2005).
In fact, they destabilize the cold gas in the galactic disk from its 
rotational equilibrium so that a fraction of it inflows toward the center (see, e.g., Cox et al. 2006). 
Such a fraction is given by the perturbation $\Delta j/j$ of the 
gas specific angular momentum $j\approx GM/v_d$ 
induced by an interaction, in the form 
\begin{equation}
f_{acc}\approx {1\over 2}\,
\Big|{\Delta j\over j}\Big|=
{1\over 2}\Big\langle {M'\over M}\,{r_d\over b}\,{v_d\over V}\Big\rangle\, ,
\end{equation}
(Cavaliere \& Vittorini 2000). Here $b$ is the impact parameter, evaluated as the average distance 
of the galaxies in the halo; $M'$ is 
the mass of the  partner galaxy in the interaction,  and the average runs over 
the probability of finding such a galaxy in the same halo where 
the galaxy $M$ is located. All such quantities are computed in our Monte Carlo SAM.{}
We calibrate the subfractions 
that go into starbursts ($f_*$) or into BH accretions ($f_{\bullet}$) 
after the limits set by  IR and X-ray surveys  of  galaxies 
with stars or  AGN  energy sources (see Alexander et al. 2005,  Franceschini et al. 2005).
We define ``stimulated" inflow the quantity $f_{acc}m_c/\Delta t=(f_*+f_{\bullet})\,m_c/\Delta t$ 
destabilized in the time step $\Delta t$; this is converted into starbursts 
and BH accretion as described below. The relevance of such stimulated inflows 
has been extensively discussed and tested by our group and others (see 
Menci et al. 2003; Cox et al. 2004; Di Matteo et al. 2005).

\noindent
{\bf Starbursts}. 
The stimulated inflow converging toward the nucleus forms starbursts at 
some $10^2 \,M_{\odot}$/yr for about $10^8$  yr (see also Somerville, Primack \& Faber 2001), 
contributing a stellar  
mass $\Delta m_* = f_*\,m_c$, with average value $\langle f_* \rangle\approx$ 15\%.  
The integrated mass so produced  is smaller than from quiescent 
formation, but it prevails in large protogalaxies at  $z > 2$. 

\noindent
{\bf Growth of massive BHs}. 
Initially, a seed BH  with  $M_{\bullet} \approx 100\,  M_{\odot}$ is assigned to each galaxy. 
At later $t$, this develops mainly by stimulated gas accretion, with average 
$\langle f_{\bullet} \rangle \approx 2$ \%. 
The result is an accretion event  of a mass $f_{\bullet}\,m_c\approx 10^{-3}\, M$ over a time close to $t_d$, 
recurring  on a time scale  $\tau$.

\noindent
{\bf AGN feedback}. 
The cold gas is re-heated also by large and impulsive energy discharges from AGNs 
activity, amounting  to $\Delta E  \approx 10^{62}\,M_{\bullet}/10^9\, M_{\odot}$ ergs.  
On coupling with the 
surrounding gas at levels $\lesssim  5\%$ (bounded by momentum conservation between photons and gas particles) these injections match 
the binding energy even in large galaxies; thus cold gas is reheated or expelled 
by superwinds and shocks (see Lapi, Cavaliere \& Menci 2005). 

\noindent
{\bf Other AGN features}. 
Our  SAM also yields  the relation $M_{\bullet}/M_{bulge}\approx 2\, 10^{-3}$  
recalled in the text, and 
quasar-AGN luminosity functions evolving in a bimodal fashion (Menci et al. 2003; 
see data by Hasinger, Miyaji \& Schmidt 2005): the bright 
early quasars peak sharply  at  $z\approx 2.5$, and drop by factors of some $10^{-3}$ 
by the present time;  the 
numerous weaker AGNs selected in X rays roll over a wide maximum around $z \approx 0.8$, 
with a weaker decline $z \approx 0$ by a factor $10^{-1}$. From the same SAM yielding these specific AGN features we 
recover the bimodal galactic colors $u-r$ shown in Fig. 2, and  predict the overall 
Eddington ratios $\lambda_E$ in Figs. 3 and 4.

\end{document}